\documentclass{elsart1p}
\usepackage{graphicx}
\usepackage{amssymb}
\begin{document}
\begin{frontmatter}
%
%
%
\title{Physics Programme of PANDA at FAIR}
%
%
\author{Inti Lehmann\thanksref{4panda}}
\ead{i.lehmann@physics.gla.ac.uk}
\thanks[4panda]{for the PANDA Collaboration}
\address{Department of Physics \& Astronomy, University of Glasgow, 
United Kingdom}
\begin{abstract}
The standard model and Quantum Chromodynamics (QCD) have undergone
rigorous tests at distances much shorter than the size of a nucleon.
Up to now, the predicted phenomena are reproduced rather well.
However, at distances comparable to the size of a nucleon, new
experimental results keep appearing which cannot be described
consistently by effective theories based on QCD. The physics of
strange and charmed quarks holds the potential to connect the two
energy domains, interpolating between the limiting scales of QCD. This
is the regime which will be explored using the future Antiproton
Annihilations at Darmstadt (PANDA) experiment at the Facility for
Antiproton and Ion Research (FAIR).

In this contribution some of the most relevant physics topics are
detailed; and the reason why PANDA is the ideal detector to study them
is given.  Precision studies of hadron formation in the charmonium
region will greatly advance our understanding of hadronic structure.
It may reveal particles beyond the two and three-quark configuration,
some of which are predicted to have exotic quantum numbers in that
mass region.  It will deepen the understanding of the charmonium
spectrum, where unpredicted states have been found recently by the
B-factories.  To date the structure of the nucleon, in terms of parton
distributions, has been mainly investigated using scattering
experiments. Complementary information will be acquired measuring
electro-magnetic final states at PANDA.
\end{abstract}
\begin{keyword}
PANDA experiment \sep antiproton \sep charmonium physics \sep exotic particles
%
\PACS 24.85.+p \sep 25.43.+t \sep 14.20.Lq \sep 14.40.Lb
\end{keyword}
\end{frontmatter}
%

The physics of strange and charmed quarks holds the potential to
connect the energy domains of perturbative Quantum Chromo Dynamics
(pQCD) and ``effective'' theories describing the properties of
nucleons, interpolating between the limiting scales of QCD. In this
regime only scarce experimental data is available, most of which has
been obtained with electromagnetic probes.

\begin{figure}[thb]
 \begin{center} \includegraphics[width=10cm]{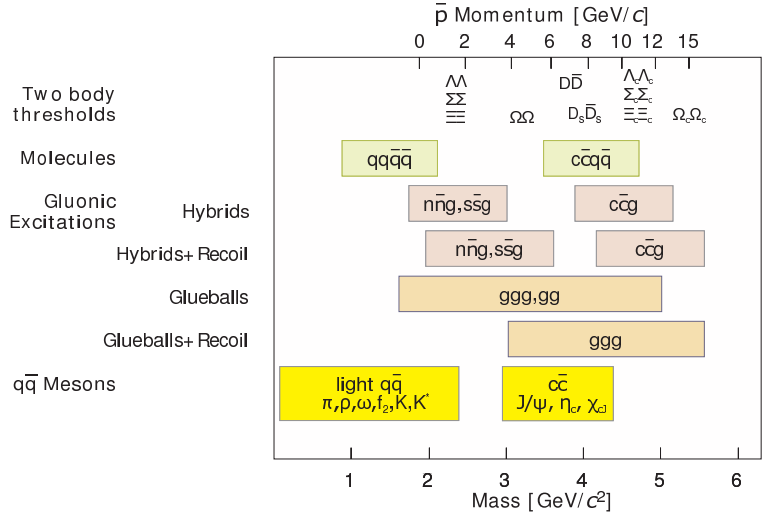}
 \caption{Mass range of hadrons that will be accessible at PANDA. The
 upper scale indicates the corresponding antiproton momenta required
 in a fixed-target experiment. The HESR will provide 1.5 to 15\,GeV/c
 antiprotons, which will allow charmonium spectroscopy, the search for
 charmed hybrids and glueballs, the production of D meson baryon pairs
 for pairs and the production of hypernuclear studies.} 
 \label{f:panda_range}
 \end{center}
\end{figure}

One possible single issue that may greatly advance our understanding
of hadronic structure is the predicted existence of states outside of
the two- and three-quark classifications, which for example could
arise from the excitation of gluonic degrees of freedom. Recent
findings from running experiments at B-factories (see {\it e.g.}
Refs~\cite{B-findings}) show that, indeed, unexpected narrow states
unaccounted for in the na\"ive quark models exist. Experiments
focussed on the abundant production and systematic studies of these
states are needed. Preferably, these should be performed using
hadronic probes because the cross sections are expected to be very
large in such systems. Results of high precision are a decisive
element to be able to identify and extract features of these exotic
states. Hadron beams are advantageous also for the production of
hadrons with non-exotic quantum numbers, as these can be formed
directly with high cross sections. Phase space cooling of the
antiproton beam furthermore allows high precision determination of the
mass and width of such states.  Using heavier nuclei as targets
enables us to investigate in-medium properties of hadrons and to
produce hypernuclei, even those containing more than one strange
quark, copiously.

The PANDA (Antiproton Annnihilation at Darmstadt) experiment (see
also Ref.~\cite{PANDA:Intro}), which
will be installed at the High Energy Storage Ring for antiprotons of
the upcoming Facility for Antiproton and Ion Research
(FAIR)~\cite{FAIR:BTR}, features a scientific programme devoted to the
following key areas.
\begin{itemize}
 \item Charmonium spectroscopy.
 \item Exotic hadrons (hybrids, glueballs, multi-quark states).
 \item Hadron properties in the nuclear medium.
 \item Strange and charmed baryons.
 \item $\gamma$-ray spectroscopy of hypernuclei.
 \item Structure of the nucleon.
\end{itemize}
Selected other topics will be studied with
unprecedented accuracy.


Conventional as well as exotic hadrons can be produced by a range of
different experimental means.  Among these, hadronic annihilation
processes, and in particular antiproton-nucleon and antiproton-nucleus
annihilations, have proven to possess all the necessary ingredients
for fruitful harvests in the hadron field.
\begin{itemize}
 \item Hadron annihilations produce a gluon-rich environment, a
       fundamental prerequisite to copiously produce gluonic
       excitations.

 \item The use of antiprotons permits to directly form all states
       with non-exotic quantum numbers (formation experiments).
       Ambiguities in the reconstruction are reduced and cross
       sections are considerably higher compared to producing
       additional particles in the final state (production
       experiments). The appearance of states in production but not in
       formation is a clear sign of exotic physics.  

 \item Narrow resonances, such as charmonium states, can be scanned
       with high precision in formation experiments using the small
       energy spread available with antiproton beams (cooled to
       $\Delta p/p = 10^{-5}$).
       
 \item Since exotic systems will appear only in production
       experiments the physics analysis of Dalitz plots becomes
       important. This requires high-statistics data samples. Thus,
       high luminosity is a key requirement. This can be achieved
       using an internal target of high density, large numbers of
       projectiles and a high count-rate capability of the detector.
       The latter is mandatory since the overall cross sections of
       hadronic reactions are large while the cross sections of
       reaction channels of interest may be quite small.  

 \item As reaction products are peaked around angles of $0^\circ$ a
       fixed-target experiment with a magnetic spectrometer is the
       ideal tool.  At the same time a $4 \pi$ coverage is mandatory
       to be able to study exclusive reactions with many decay
       particles.  The physics topics as summarised in
       Fig.~\ref{f:panda_range} confirm that the momentum range of
       the antiproton beam should extend up to 15\,GeV/c with
       luminosities in the order of $10^{32}\,$cm$^{-2}$s$^{-1}$
\end{itemize}

The PANDA collaboration is prepared to address these topics with a
general-purpose internal-target experiment utilising the antiprotons
provided at the upcoming FAIR facility (see also
Refs.~\cite{Lars:PANIC08,PANDA:Intro}). Within the growing PANDA
collaboration of currently about 400 physicists from 16 countries, an
extensive R\&D programme is under way, which comprises already a
detailed design of the detector. PANDA gratefully acknowledges the
support of the respective national research agencies and the European
Union funds.

%
%
%

%
\end{document}